\documentstyle[preprint,aps]{revtex}

\def\be{\begin{equation}}
\def\en{\end{equation}}
\def\bq{\begin{eqnarray}}
\def\eq{\end{eqnarray}}

\begin{document}

\begin{center}
{\Large {\bf On Heat Flow and Non-Reciprocity}}\\[1.5cm]
\end{center}


\begin{center}
{\large {\bf Alejandro Cabo Montes de Oca}}\\{\it Group of Theoretical
Physics\\Instituto de Cibern\'{e}tica Matem\'{a}tica y F\'{\i}sica}\\{\it %
Calle E, No. 309. Esq. a15, Vedado, La Habana, Cuba}
\end{center}

\noindent

\vspace{0.1cm}

\begin{center}
{\bf Abstract.}

\noindent
Heat transfer properties in non-reciprocal systems are discussed. An ideal
experiment employing microwave or optical isolators is considered in order
to investigate the possibility for a spontaneous transfer of energy between
two black-bodies at common temperatures. Under the validity of the adopted
assumptions on the examined physical system, the effect appears to be weak
but feasible.

\end{center}

\newpage

\section{Introduction.}

The second law of thermodynamics is one of the basic principles of this
discipline\cite{zeman}. Its foundation from the point of view of the more
basic branch : Statistical Physics is well established and its validity is
recognized for an enormous range of physical systems and phenomena \cite
{landau} . However, it is also admitted that the definition of Entropy for
systems staying far away from thermal equilibrium is not obvious \cite
{ruelle}. Therefore, the theoretical proof of the fact that Entropy is
always growing in non-equilibrium processes (occurring in a completely
arbitrary isolated physical system) is also not fully established. This
situation confers an status of a basic but heuristic principle to the Second
Law. In addition, it is also certain that the standard derivations of the
Law from Statistical Physics assume conditions which are obeyed by an
enormous class of physical systems but which, in spite of it, are by far non
general and excludes less common but very important phenomena in Nature. One
particular requirement, which eliminates a large class of situations, is the
condition that the spectrum of the Gibbs subsystems are statistically
independent among them\cite{landau}. This property is not satisfied when
long range forces are present. Such kind of limitations have given ground to
the work of C. Tsallis \cite{tsallis} on introducing alternative expressions
for the Entropy for long range interacting systems. This approach recently
have received an experimental support \cite{tsallis1} . However, although
new expressions for the entropy are under consideration the general
expectation is the survival of the Law of increasing Entropy for isolated
systems within a more general picture.

In this work we intend to attract the attention about another aspect of the
derivation of the $2^{nd}$ Law from Statistical Physics. It is related with
the physics of the heat transfer among bodies and the fact that such energy
flows are expected to always spontaneously occur from higher to lower
temperature regions. This principle is obtained form the Law of increasing
Entropy and also implies it. However, its theoretical derivation from
statistical physics has received very much limited attention. It can be
understood that for wide, but even though, particular class of physical
systems having reciprocal dynamics, the heat will always travel from higher
to lower temperature regions. When, however, the system has a non-reciprocal
dynamical response this same property is less intuitively evident.


Let us examine for example the following situation. Consider the statistical
mean value of the Poynting vector operator, reflecting the electromagnetic
energy flux in a many body system coupled with the electromagnetic field. It
will be assumed also that the system is subject to an homogeneous and
constant magnetic field which therefore destroys the exact parity
invariances of the photon propagator and then its reciprocity property. The
general expression for the mean value of the Poynting vector operator can be
shown to vanish under the assumed translation invariance of the system.
However, it can be underlined that in the non reciprocal situation the
vanishing of the net energy flux follows from a dynamical equilibrium in
which the energy flux of the modes propagating in exact opposite directions
exactly cancels, but their polarization vectors does not show identical
spatial distribution propagating in opposite senses. This property suggests
the possibility that a coupling of such systems with bodies showing
reciprocal dynamics could lead to an unbalanced flux of the Poynting vector.

This phenomenon effectively happens in devices such as optical and microwave
isolators \cite{hand1},\cite{hand2},\cite{hand3}. However, up to our
knowledge, the analysis of the implications of such properties for the
energy transfer at thermal equilibrium is not a fully investigated and
understood matter.

This work intends to discuss this question. For that purpose an analysis of
two types of isolators connecting two regions containing blackbody radiation
at common temperatures is done. Within the adopted assumptions, the results
seem to indicate the possibility for the occurrence of a stationary energy
flow between the two blackbody regions. Although the magnitude of the fluxes
evaluated are far from being of practical relevance, the correctness of the
analysis could support the search of similar effects in which larger energy
flows could occur.

In the next Section 2, concrete versions of the isolator devices to be
discussed are presented and the power input at entries are evaluated. In
Section 3, the energy flow balance is investigated by comparing the
situations in which the non reciprocity is switched on and off.

\section{Optical and Microwave Isolators}

Let us consider a system composed by two big reservoirs $S\_$ and $S_{+}$
which are thermally isolated between them by a wall $W$ \ having a low value
of the thermal conductivity $\sigma $ due collisions. The wall is also
assumed to be highly absorptive for electromagnetic radiation in a wide
frequency region containing most of the black-body spectrum at the
temperature value $T\ $ to be considered. The system is illustrated in
Figs.1.

The wall $W$ is considered as permeated by a square lattice of special non
reciprocal devices called isolators \cite{lax}, separated one from another
by a lattice period $d$ . Let us consider in what follows two types of such
devices. The so called resonance isolators which are based in the effect of
non reciprocal resonance absorption in ferrites; and the other ones which
operation is determined by the Faraday rotation in the same kind of
materials. For resonance isolators the width $h$ of $W$ is assumed to
coincide with the length $2l+l_1$ of the isolators , where $l_1$ is the
length of a thin ferrite rod furnishing the non reciprocal propagation
properties of the device. The symbol $l$ indicates the length of two
unloaded rectangular waveguides serving as the entries for the portion
loaded with the ferrite rod. The structure is shown in Fig. 2.

In the case of the Faraday effect isolator, the width of $W\ $is also $%
h=2l+l_1$ where again $l$ is the length of the rectangular waveguide
entries, but $l_1\ $is now the length of a cylindrical waveguide at which
axis a ferrite rod producing the Faraday rotation of the waves is situated.
In addition, the output rectangular waveguides have their corresponding
sides rotated in $\frac \pi 4$ radians between them. Also, at one of the
entries there is a thin resistive vane with its plane being parallel to the
wide side of the rectangular waveguide. The structure is illustrated in Fig.
3. More details on the operation of such devices can be found in \cite{hand1}%
,\cite{hand2},\cite{hand3}.

It can be remarked that such isolators have been extensively studied and
applied in microwave technology. The resonance devices have been
investigated theoretically by B. Lax \cite{lax}. In general they can be
developed in a way showing, by example, 60 db of attenuation for propagation
in one sense and allowing the power in the other sense to have as low as 0.5
db of losses.

The general aim of the work is to investigate the flux of heat flowing
through each isolator in the wall $W$ . The region at the left of $W$ will
be designated by $S_{-}\ $and the one at the right by $S_{+}.$ The
black-body radiation filling them will be assumed to have a common value of
the temperature $T_{}.$ Calling $a$ and $b$ the longer and shorter sides of
the rectangular waveguide entries, the dispersion relations for the EM modes
of propagation in these input guides are given by

\begin{equation}
\frac wc=\frac \epsilon {c\hbar }=\sqrt{\frac{(m\pi )^2}{a^2}^{}+\frac{(n\pi
)^2}{b^2}+k^2}=\sqrt{(\frac{\epsilon _c^{mn}}{c\hbar })^2+k^2},
\end{equation}

\begin{equation}
\frac{w_c^{mn}}c=\frac{\epsilon _c^{mn}}{c\hbar }=\sqrt{\frac{(m\pi )^2}{a^2}%
^{}+\frac{(n\pi )^2}{b^2}},
\end{equation}

\noindent
where $m$ and $n$ are any positive integers, $k=\frac{2\pi }\lambda \ $is
the wavevector associated to the wavelength $\lambda $ in the guide and the $%
w_c^{mn}$ and $\epsilon _c^{mn}$ are the cutoff frequencies or energies for
the $(m,n)\!\ $propagation modes. Relation (1) defines the photon energy $%
\epsilon $ within the waveguide. The lowest cutoff frequency is given by

\begin{equation}
\frac{w_c^{10}}c=\frac \pi a.
\end{equation}
Assuming $a=2\ b$ the next cutoff frequency turns to be the double of the
lowest one:

\begin{equation}
\frac{w_c^{01}}c=\frac{2\pi }a.
\end{equation}

Then, the density of states per unit energy $\rho (\epsilon )\ $of the
lowest mode vanishes for values of the energy lower than $\epsilon _c\ $and
above this value it is given by the expression 
\begin{eqnarray}
dN(\epsilon ) &=&\frac{\Delta k}{\frac{2\pi }L}=\frac{dk}{d\epsilon }\frac{%
d\epsilon }{\frac{2\pi }L} \\
&=&\frac L{2\pi \hbar c}\frac{\epsilon \ d\epsilon }{\sqrt{\epsilon
^2-(\epsilon _c^{10})^2}}  \nonumber \\
&=&\rho (\epsilon )\ d\epsilon .  \nonumber \\
&&  \nonumber
\end{eqnarray}
Therefore, after multiplying by the Bose distribution function and the
photon energy $\epsilon ,$ the amount of thermal energy per unit photon
energy propagating in one of the two allowed senses is given by 
\begin{eqnarray}
\frac{du(\epsilon )}{d\epsilon } &=&\rho (\epsilon )\ \epsilon \ \
f(\epsilon ), \\
\ f(\epsilon ) &=&\frac 1{\exp (\frac \epsilon {kT})-1},  \nonumber
\end{eqnarray}

\noindent
where $k$ and $T\ $are the Boltzman constant and the temperature
respectively. The power per unit photon energy transmitted in any of the
senses in the entry guides is then given by

\begin{equation}
dP(\epsilon )=\frac 1{T_{trans}}\frac{du(\epsilon )}{d\epsilon }d\epsilon ,
\end{equation}

\noindent
where the transit time $T_{trans}\ $through a length $L$ of the rectangular
waveguide for a photon of energy $\epsilon ,$ can be obtained though the
following argument. Let us consider the fact that the propagation of the
first mode can be interpreted as the total reflection of two plane waves
which are reflected by two parallel metallic planes obtained by extending
the shorter sides of the waveguides. These waves have a momentum component
along the waveguide given by $k$ and a component along the large side given
by $\pm \frac \pi a.$ Hence, the propagation along the waveguide axis has
the velocity 
\begin{eqnarray}
v_{\parallel } &=&\frac{c\ k}{\sqrt{\frac{(\pi )^2}{a^2}+k^2}} \\
&=&c\ \frac{\sqrt{\epsilon ^2-(\epsilon _c^{10})^2}}\epsilon .  \nonumber
\end{eqnarray}

Thus, the time of transit through a length $L$ of the guide and the density
of power per unit energy in the entry guides are given by 
\begin{equation}
T_{trans}=\frac L{v_{\parallel }}=\frac Lc\frac \epsilon {\sqrt{\epsilon
^2-(\epsilon _c^{10})^2}}.
\end{equation}

\begin{eqnarray}
dP(\epsilon ) &=&\frac 1{T_{trans}}\frac{du(\epsilon )}{d\epsilon }d\epsilon
\\
&=&\frac 1{2\pi \hbar }\frac \epsilon {\exp (\frac \epsilon {kT})-1}=\frac{%
dP(\epsilon )}{d\epsilon }d\epsilon .  \nonumber
\end{eqnarray}

Let us discuss now the connections between the dynamical and thermodynamical
properties in the system under consideration. We would like to stress that
the above considered power flow at the entry guides should be established
dynamically by the incidence of the black body radiation at both sides of
the wall $W.$ Let us assume that the transmission coefficient of the entry
guides are near to unity at the considered frequency region. The same will
be assumed to happens for the rectangular or cylindrical waveguides when the
ferrite material is absent. This condition is affordable in transmission
lines from microwave to optical regions at laboratory values of the
waveguide sizes\cite{montgomery}. In other word, it will be assumed that the
possible skin or radiation losses at the waveguides is low in order that the
decay length of the mode power is much greater than the isolator dimensions.

Moreover, in both cases of the resonance and Faraday rotation isolators, it
will be assumed that the ferrite kernel is saturated. This will allow to
employ the known expressions for the magnetic susceptibilities in such cases
in order to estimate the energy flow in that structures.

\section{ Reciprocity and Energy Flow Balance}

Let us analyze now the radiation which is transmitted by the isolators
between the regions $S_{-}$and $S_{+}.$ For this purpose, consider that a
control device is used to switch on the ferrite in its active positions for
the resonance or Faraday rotation type isolators. Before the switching on,
the system is completely reciprocal. Since we are assuming that the
emittance and absorptivity of the walls of the waveguides are small
(transmission coefficients near to unity), the power at their interior
should be generated mainly by the black body radiation incident at both
sides of the wall in each isolator (whenever the inputs are not blocked). If
the inputs are poorly matched or closed, then the small but non vanishing
emittance of the walls should be the main dynamical source of the thermal
power spectrum at the inside. However, when the inputs are nearly matched ,
the energy of that sources should be insufficient to elevate the power
spectrum to the level given by Eq.(10). Therefore, within the adopted
conditions, it seems possible to assume that the incident power at the entry
waveguides is propagating as coming from the input and not stochastically
generated at the interior. This is a main assumption in the present
analysis. In the absence of the ferrite, the vanishing of the net power
transmitted from $S_{-}\ $to $S_{+}\ $can be easy understood. For example,
in the resonance isolator what connects both regions is a piece of
rectangular waveguide of total length $h.$ Let us consider two imaginary
surfaces covering the inputs and being completely symmetrical under a
reflection in a plane parallel to $W$ and equidistant from both sides. Also
assuming the boundary conditions of the electromagnetic radiation as
represented by sources (through the use of the Green Theorem ) the
reciprocity property assures that, since the statistical distribution of the
sources representing the blackbody radiation are identical in both surfaces,
the power propagating in both senses should be identical. It should be
noticed that this argument also implies that when the temperature of the
blackbody radiation is higher at one of the sides, let say $S_{-\text{,}}$
the energy (heat) should flow from the high to lower temperature regions
because the equivalent sources are stronger at the higher temperature side.

In the case of the Faraday isolator the vanishing of the net transmitted
power can also be understood dynamically. This is because the polarization
of the waves coming from $S_{-}$ to $S_{+}$ have a $\frac \pi 4\ $rotation
with respect to the small side of the of the external waveguide and then
only a fraction of the power will be transmitted after multiple
reflections.. Equivalently, the wave coming from $S_{+}$ to $S_{-}$will also
arrive with a $\frac \pi 4\ $angle of polarization with the short side of
the entering waveguide at $S_{-}$. Then, as the solutions of the
electromagnetic propagation are symmetrical under reflection in the above
mentioned symmetry plane, it follows that the fraction of the power
transmitted in this case should be identical as in the former situation.
This remarks assure that the net energy flux from one side to another
exactly vanishes.

In the presence of the magnetic effects, the exact reflection invariance is
absent and the above arguments can't be applied.

Let us consider below the switching on of the ferrite in either of the both
types of isolators.

\subsection{\noindent
Faraday Isolator.}

After putting the ferrite in its working position, the amount of incident
power given by Eq. (10) and flowing from the $S_{-}$ should propagate with
relative small losses (as assumed before) up to the other output whenever
the Faraday rotation angle is fixed to be near $\frac \pi 4$ in the
appropriate sense. This situation can be attained by commuting the sense of
the saturation magnetization direction. Then, a large fraction of the power
given by expression (10) can be transmitted to the other side of the
isolator if the coupling between the cylindrical and rectangular waveguides
is near to be matched. In another hand, the radiation propagating from the
entry waveguide at $S_{+}$ in direction of $S_{-\text{ }}$will suffer almost
a total reflection at the junction between the cylindrical to rectangular
waveguide near $S_{-\text{ }},\ $since the non reciprocal dielectric action
of the ferrite will rotate an additional $\frac \pi 4\ $the polarization of
the electric field. This fact makes incompatible the symmetry of the wave
with the excitation of the lowest mode in the output rectangular waveguide.
After the total reflection and the also propagation back the energy of this
wave is dissipated at the resistive vane $R\ $at the $S_{+}$side in Fig. 3.

The dissipation at the resistive plate should be discussed in more detail.
The power absorbed by the vane can be partially radiated back and partially
thermally conducted to the $S_{+}\ $side by a good thermal matching of the
ferrite with the waveguide walls. Let us consider a design of the resistive
card in which the excess power dissipated in it can be efficiently
transferred to the waveguide walls and from them to the $S_{+}.\ $This
procedure should allow to reduce the amount of the dissipated power returned
back as non polarized radiation to the waveguide. However, even in the case
that the full power absorbed is radiated back in the waveguide (a very
improbable outcome) only a fraction of it will have the appropriate
polarization to emerge at the $S_{-}\ $side. Thus, even in this extreme
case, the balance of the energy fluxes is not evidently occurring.

Let us consider now the determination of the amount of power which could be
transmitted if the argued mechanism is effectively at work. This quantity
can be evaluated in the following way. Assume that the Faraday angle is
adjusted to be $\frac \pi 4\ $for a particular value of the frequency , for
example $\sqrt{2}w_c^{10}\ $lying below the cutoff of the second propagation
mode. Then, also suppose the working frequencies$\ w\ $are so high that the
magnetic Faraday susceptibility show its far from resonance asymptotic
dependence\cite{hand2} :

\begin{equation}
\chi _F\sim \frac{w_M}w,
\end{equation}

\noindent
where $w_M\ $is proportional to the ferrite magnetization. As the Faraday
rotation is proportional to $\chi _F$, the rotation angle as a function of
the energy $\epsilon \ $(or frequency $w$) can be evaluated.

Therefore, the fraction of the power within some energy interval $d\epsilon
, $ which will transmitted from $S_{-}$to $S_{+}$ will be given by the
squared cosine of the difference between the Faraday rotation angle and $%
\frac \pi 4.\ $ The rest of the power is dissipated in the resistive vane $R$
at the entering of the cylindrical guide near $S_{+}.\ $ In another hand,
the wave incident in the cylindrical guide trough $S_{+}\ $is rotated $\frac 
\pi 4\ $plus the Faraday angle with respect to the small axis of the squared
waveguide at $S_{-\text{ }}$. Then, the power transmitted to the matched
rectangular waveguide entering at $S_{-}\ $is the fraction given by the
squared cosine of that angle. The rest of this power is reflected back and
the fraction able to pass into the rectangular waveguide escapes, and the
other portion is dissipated at the resistive load. Finally, as the incoming
powers at the rectangular waveguides are equal, the net energy flux (the
integral over all the energies) can be estimated to be given by:\ 

\begin{equation}
P=\frac{\epsilon _c^{10}}{2\pi \hbar }\int_1^\infty dx\ x\frac 1{\exp (\frac{%
\epsilon _c^{10}x}{kT})-1}\left( \cos ^2(\frac \pi 4(\frac{\sqrt{2}}x%
-1))-\sin ^2(\frac \pi 4(\frac{\sqrt{2}}x-1))\right) .
\end{equation}

\noindent
The integral is only dependent on$\ \frac{\epsilon _c^{10}}{kT}.$ For the
case $\frac{\epsilon _c^{10}}{kT}=1$ the expected power has the expression

\begin{equation}
P=0.7018\ \frac{k^2T^2}{2\pi \hbar },
\end{equation}
which at the room temperature $T=300\ K^o,\ $predicts a net transmitted
power of 
\[
P=0.18149\ \ erg/s. 
\]

As it could be expected form the radiative nature of the discussed
processes, this a small power transfer. But, it also should be underlined,
that at room temperature the size of the waveguide satisfying the condition $%
\frac{\epsilon _c^{10}}{kT}=1$ is as small as $24\ \mu m.$ Then, after
assuming that the lattice of isolators has $d=50\ \mu m\ $as the size of the
square unit cell, the power per squared meter can rise up to near $7$ $watt\
/\ m^2.$ Even in this case the considered heat flux becomes of no
significance for practical purposes.

However, with respect to the detectability the situation seems to be
different. After assuming that the estimated power in a single isolator Eq.
(13) is dissipated in a let say standard resistance of $75\ \Omega ,$ the
voltage produced should be of the order of $1\ mV.$ \ This result indicates
the possibility for the detection of the proposed effect under the current
experimental conditions.

\subsection{\noindent
 Resonance Isolator.}

Next let us consider the switching of the ferrite rod in the resonance
isolator. The following conditions will be also assumed:

1) The ferrite will be supposed to have an efficient heat coupling with the
metallic sides of the rectangular waveguides, thus assuring that the power
dissipated in it is transferred in a large proportion to the cavity walls.

2) At the same time, the cavity walls are assumed to have a low thermal
resistance for the conduction towards the $S_{+}\ $side and a very much
higher one for the conduction into the $S_{-}\ $region.

I think that these could not be essentially difficult conditions to match.
They basically depend on seemingly controllable aspects of design for the
devices, at least, whenever an ideal situation with respect to the available
values of the heat conductivities is assumed.

After the switching on of the ferrite the wave can excite the ferromagnetic
resonance of the material. The resonance frequency will be assumed to have a
value between $w_c^{10}$and $w_c^{01}.$ It can be stressed that at points
being near the small side of the waveguides a distance $\frac a4,$ the
microwave magnetic field is circularly polarized. This is a typical position
selected for the ferrite rod.

However, the resonance occurs only for one of the senses of rotation for the
polarization vector. Since the wave going in opposite spatial senses in the
first mode have opposite senses of rotation for their circular polarization,
it occurs that one wave can be absorbed much more than the other. Under the
change in sign of the magnetization the behavior of the two waves are
interchanged. Let us assume that the magnetization is polarized in the sense
that produces resonance for the waves going from $S_{+}$ to $S_{-}$ . Then,
when the ferrite is connected, that wave becomes absorbed and the reverse
one is less disturbed. Then, if the absorbed power in the ferrite is
transferred efficiently to the waveguide walls and from them to the $S_{+}$
side, the amount of power passing from $S_{-}$ to $S_{+}$ appears to be
higher that the one flowing in the reverse sense. It should be noticed that
for this conclusion to be valid it is essential to assume that the
emittances at thermal equilibrium of the waveguide walls are very low. Then,
it seems improbable that their effect will compensate for the absorption of
the wave coming from $S_{+}.$\ 

Let us evaluate below the net transmitted power which could be expected .
For this purpose consider the transmission coefficients for the waves
propagating in both senses. As cited before, the propagation in waveguides
loaded with ferrite bars, as depicted in Fig. 2, have been studied
theoretically by Lax \cite{lax}, \cite{lax1} . The analysis can be done
using the perturbative approximation for the derivation of the propagation
modes starting from the unperturbed ones \cite{hand1}.

The transmission coefficients for a length $l_{1\text{ }}$loaded with a thin
bar of ferrite as in Fig. 2 are given in the above mentioned perturbative
approximation, by

\begin{eqnarray}
T_{-+} &=&\exp (-2Re\left[ \Gamma _{-+}\right] \ l_1), \\
T_{+-} &=&\exp (-2Re\left[ \Gamma _{+-}\right] \ l_1),  \nonumber \\
&&  \nonumber
\end{eqnarray}
where the absorptive parameters $\Gamma _{-+}\ $and $\Gamma _{+-}$ are
defined by the expressions\cite{hand1} 
\begin{equation}
\Gamma _{-+}=\frac{\Delta S}S\frac 1k\left[ \left( k^2\sin ^2\left( \frac{%
\pi x}a\right) +\left( \frac \pi a\right) ^2\cos ^2\left( \frac{\pi x}a%
\right) \right) \chi _{xx}-2\ \chi _{xy}\sin \left( \frac{\pi x}a\right)
\cos \left( \frac{\pi x}a\right) \right] ,
\end{equation}
$_{}$%
\begin{equation}
\Gamma _{+-}=\frac{\Delta S}S\frac 1k\left[ \left( k^2\sin ^2\left( \frac{%
\pi x}a\right) +\left( \frac \pi a\right) ^2\cos ^2\left( \frac{\pi x}a%
\right) \right) \chi _{xx}+2\ \chi _{xy}\sin \left( \frac{\pi x}a\right)
\cos \left( \frac{\pi x}a\right) \right] ,
\end{equation}
in which it is assumed that$\ $the non diagonal component of the magnetic
susceptibility satisfies $\chi _{xy}>0;$ $S,\Delta S\ $ are the areas of the
sections of the waveguide and the ferrite bar respectively and $x$ is the
distance of the center of bar from the small side of the waveguide. The
diagonal and non diagonal components of the magnetic susceptibility are
given by \cite{hand2}

$_{}$%
\begin{eqnarray}
\chi _{xx} &=&\frac{\omega _M\left( \omega _0+j\omega _L\right) }{\left(
\omega _{0+}j\omega _L\right) ^2-\omega ^2}, \\
\chi _{xy} &=&\frac{\omega _M\omega j}{\left( \omega _{0+}j\omega _L\right)
^2-\omega ^2}.  \nonumber
\end{eqnarray}
The various frequency parameters in (17) are defined as 
\begin{eqnarray}
\omega _0 &=&2\pi \gamma H_0, \\
\omega _M &=&2\pi \gamma \left( 4\pi M_s\right) \ \mu _0,  \nonumber \\
\omega _L &=&\frac 1T=\frac{2\pi \gamma \Delta H}2,  \nonumber \\
4\pi M_s &=&saturation\ magnetization,  \nonumber \\
T &=&macroscopic\ relaxation\ time,  \nonumber \\
\gamma &=&2.8\ MHz/Oe.  \nonumber
\end{eqnarray}

Taking the real part of the $\Gamma \ $parameters, the transmission
coefficients can be written as

\begin{eqnarray}
T_{-+}(\epsilon ) &=&\exp (-\frac{2l_1\Delta S}S\frac{\epsilon _m\epsilon _L%
}{\left( \epsilon ^{2-}\epsilon _0^2+\epsilon _L^2\right) ^2+4\epsilon
_L^2\epsilon _0^2}\frac{c\hbar }{\sqrt{\epsilon ^2-(\epsilon _c^{10})^2}} \\
&&(\left( \frac{\epsilon ^2-(\epsilon _c^{10})^2}{c^2\hbar ^2}\sin ^2\left( 
\frac{\pi x}a\right) +\left( \frac \pi a\right) ^2\cos ^2\left( \frac{\pi x}a%
\right) \right) (\epsilon _0^2+\epsilon _L^2+\epsilon _{}^2)-  \nonumber \\
&&4\sin \left( \frac{\pi x}a\right) \cos \left( \frac{\pi x}a\right) \frac 
\pi {a\hbar c}\sqrt{\epsilon ^2-(\epsilon _c^{10})^2}\epsilon _0\epsilon )),
\nonumber
\end{eqnarray}

\begin{eqnarray}
T_{+-}(\epsilon ) &=&\exp (-\frac{2l_1\Delta S}S\frac{\epsilon _m\epsilon _L%
}{\left( \epsilon ^2-\epsilon _0^2+\epsilon _L^2\right) ^2+4\epsilon
_L^2\epsilon _0^2}\frac{c\hbar }{\sqrt{\epsilon ^2-(\epsilon _c^{10})^2}} \\
&&(\left( \frac{\epsilon ^2-(\epsilon _c^{10})^2}{c^2\hbar ^2}\sin ^2\left( 
\frac{\pi x}a\right) +\left( \frac \pi a\right) ^2\cos ^2\left( \frac{\pi x}a%
\right) \right) (\epsilon _0^2+\epsilon _L^2+\epsilon _{}^2)+  \nonumber \\
&&4\sin \left( \frac{\pi x}a\right) \cos \left( \frac{\pi x}a\right) \frac 
\pi {a\hbar c}\sqrt{\epsilon ^2-(\epsilon _c^{10})^2}\epsilon _0\epsilon )),
\nonumber
\end{eqnarray}
where the newly used energy parameters are defined as

\begin{eqnarray}
\epsilon _0 &=&\hbar \ w_0, \\
\epsilon _m &=&\hbar \ w_m,  \nonumber \\
\epsilon _L &=&\hbar \ w_L.  \nonumber
\end{eqnarray}

It is possible to write now for the total transmitted power the expression

\begin{eqnarray}
P=\int_{\epsilon _c}^\infty d\epsilon \frac{dP(\epsilon )}{d\epsilon }%
(T_{-+}\left( \epsilon \right) -T_{+-}\left( \epsilon \right)
)=\int_{\epsilon _c}^\infty \frac 1{2\pi \hbar }\frac{\epsilon \ d\epsilon }{%
\exp \left( \frac \epsilon {kT}\right) -1}(T_{-+}\left( \epsilon \right)
-T_{+-}\left( \epsilon \right) ) &&. \\
&&  \nonumber
\end{eqnarray}
Let us assume that $x=\frac a4,$ as the typical value considered for
resonance isolators. Then the formula for the power can be rewritten as
follows 
\begin{eqnarray}
P &=&\frac{\epsilon _c^2}{2\pi \hbar }\int_1^\infty dx\ x\sum_{\sigma =\pm }%
\frac \sigma {\exp \left( \frac{\epsilon _cx}{kT}\right) -1}\times \\
&&\exp \left( -\frac{l_1\Delta S}{2S}\frac{\epsilon _m\epsilon _L}{\hbar c\
\epsilon _c}\frac{\left[ x^2\left( x^2+\left( \epsilon _0/\epsilon _c\right)
^2+\left( \epsilon _L/\epsilon _c\right) ^2\right) -4\sigma \ x\ \epsilon _0%
\sqrt{x^2-1}/\ \epsilon _c\right] }{\sqrt{x^2-1}\left( \left[ x^2-\left(
\epsilon _0/\epsilon _c\right) ^2+\left( \epsilon _L/\epsilon _c\right)
^2\right] ^2+4\epsilon _L^2\epsilon _0^2/\epsilon _c^4\right) }\right) 
\nonumber
\end{eqnarray}
where $\epsilon _c^{}=$ $\epsilon _c^{10}$ is the cutoff energy of the first
mode.

Let us select specific values for the parameters in order to get sense of
the order of magnitude for the expected powers which could be attained.
Suppose also that the ferromagnetic resonance frequency is given by $w_0=%
\sqrt{2}\ \frac{\epsilon _c}\hbar ,\ $which means that the resonance occurs
at $\sqrt{2}\ $times the cutoff frequency of the first mode and also that
the resonance width takes the value $w_L=0.2\ \frac{\epsilon _c}\hbar .$
Moreover, let us fix the other parameters through adjusting the following
relation to be valid 
\begin{equation}
\frac{l_1\Delta S}{2S}\frac{\epsilon _m\epsilon _L}{\hbar c\ \epsilon _c}%
=0.1.
\end{equation}

This condition assures a relatively low absorption for the power of the
waves travelling from $S_{-}$to $S_{+}.$ Finally, let fix again the
condition that the dimension of the waveguide is such that the thermal
energy satisfies

\begin{equation}
kT=\epsilon _c.
\end{equation}
After that, the numerical integration of the relation (23) gives the result 
\begin{eqnarray}
P &=&0.36320\ \frac{(kT)^2}{2\pi \hbar }, \\
&=&0.36320\ \frac{\epsilon _c{}^2}{2\pi \hbar }=0.36320\ \frac{(\pi
c/a){}^2\hbar }{2\pi }=\!\frac{1.47\times 10^{-7}}{a^2}\ erg/s,  \nonumber \\
&&  \nonumber
\end{eqnarray}
in which the values for $\hbar =1.054572\ 10^{-27}erg.s\ $and $c=3\ 10^{10}\
cm/s\ $have been substituted. It can be observed that for waveguides of
sizes of the order of a centimeter the output power is very low. At the
ambient temperature $T=300\ K^o,$ which is equivalent to a waveguide
dimension $a$ of the order of $24\ \mu m,$ the power transmitted by one
isolator should be 
\[
P=0.02552\ erg/s, 
\]
which is of the same order of magnitude of the one estimated for the Faraday
rotation situation. Henceforth, it became clear, that in both of the
considered cases, the implied powers are low. However, the real existence of
such an effect could open the possibility for the search of alternative
kinetic processes which would show improved power fluxes. However, in the
sense of the possibility of detecting the effect, the results give a
positive answer, at least in the cases of infrared or optical isolators. For
them, the signals to be detected stay within the voltage ranges of the
standard voltage or electric current measurement instruments. 
\[
\]

\noindent {\large {\bf Acknowledgments}}

I would like to acknowledge the helpful discussions with Drs. H. Wio, A.
Gutierrez and R. Franco; and the support during the development of this work
of the Associate Program of the Abdus Salam International Centre for
Theoretical Physics and the Theoretical and Experimental Divisions of CERN.%
\newpage

{\bf Figure Captions }

{\bf Fig1}. Scheme of the system under consideration. The $S_{-}$ and $S_{+}$
are regions each of them containing black-body radiation at a common
temperature $T.\ ${\bf \ }

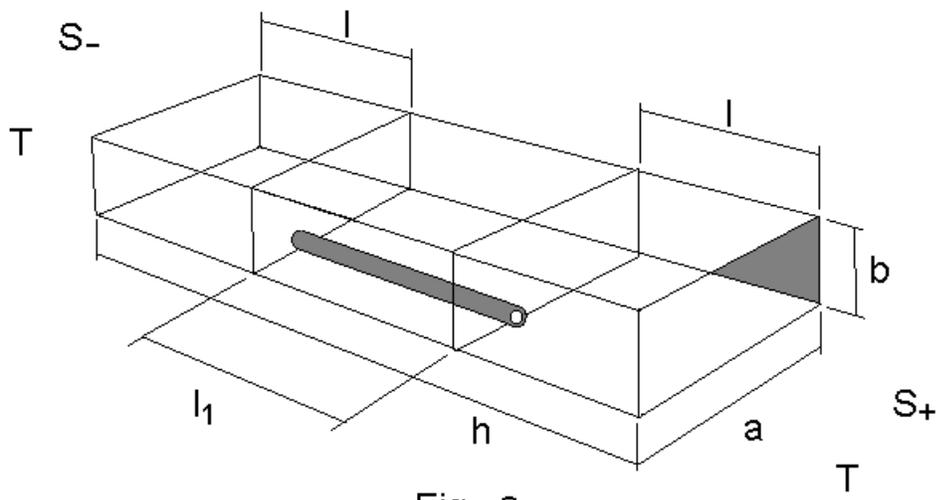
{\bf Fig.2}. Picture of the type of resonance isolator considered in this
work.

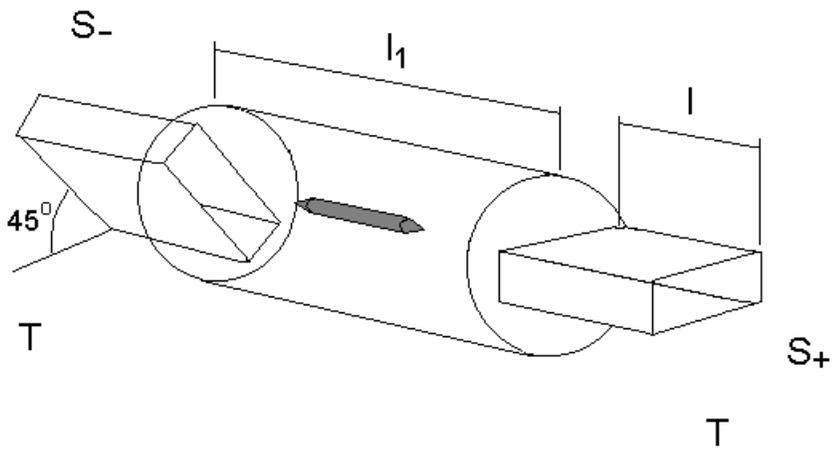
{\bf Fig.3}. Picture of the particular type of Faraday rotation isolator
discussed in the text.

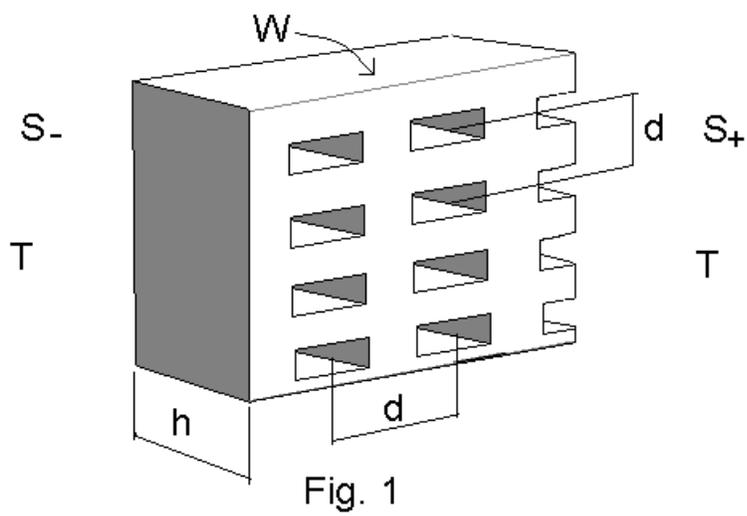
\end{document}